\documentclass[a4paper,11pt]{article}
\pdfoutput=1 

\usepackage{jcappub} 
\usepackage{xcolor}
\usepackage[T1]{fontenc} 
\usepackage{bm}
\usepackage{epsfig}
\usepackage[lofdepth,lotdepth,caption=false]{subfig}
\usepackage{graphicx}

\usepackage{float}
\usepackage{graphicx}
\usepackage{amsmath,amssymb}
\usepackage{textcomp}
\usepackage{gensymb}
\usepackage[utf8]{inputenc}
\usepackage[T1]{fontenc}
\usepackage{setspace}
\usepackage{hyperref}
\usepackage{color}
\usepackage{slashbox}
\usepackage{bbold}

\def\be{\begin{equation}}
\def\ee{\end{equation}}
\def\bea{\begin{eqnarray}}
\def\eea{\end{eqnarray}}

\begin{document}

\title{Explaining the ANITA events by a $L_e-L_\tau$ gauge model}

\author[a,c]{Arman Esmaili,}
\author[b,c]{Yasaman Farzan}
\emailAdd{arman@puc-rio.br}
\emailAdd{yasaman@theory.ipm.ac.ir}
\affiliation[a]{Departamento de F\'isica, Pontif\'icia Universidade Cat\'olica do Rio de Janeiro, Rio de Janeiro 22452-970, Brazil}
\affiliation[b]{School of physics, Institute for Research in Fundamental Sciences (IPM), P.O. Box 19395-5531, Tehran, Iran}
\affiliation[c]{The Abdus Salam ICTP, Strada Costiera 11, 34151, Trieste, Italy}

\abstract{
The ANITA experiment has registered two anomalous events that can be interpreted as $\nu_\tau$ or $\bar{\nu}_\tau$ with a very high energy of $\mathcal{O}(0.6)$~EeV emerging from deep inside the Earth. At such high energies, the Earth is opaque to neutrinos so the emergence of these neutrinos at such large zenith angles is a mystery. In our paper, we present a model that explains the two anomalous events through a $L_e -L_\tau$ gauge interaction involving two new Weyl fermions charged under the new gauge symmetry. We find that, as a bonus of the model, the lighter Weyl fermion can be a dark matter component. We discuss how the ANITA observation can be reconciled with the IceCube and Auger upper bounds. We also demonstrate how this model can be tested in future by collider experiments.}

\maketitle
\date{\today}


\section{Introduction}

Various observatories taking data on different ranges of the electromagnetic radiation, the current and planned gravitational wave detectors, cosmic ray detectors and, last but not least, neutrino telescopes have together ushered in the multimessenger era for exploring the cosmos as well as for learning about possible exotic properties of the elementary particles. One of the key players in this scene is ANITA (Antarctic Impulse Transient Antenna)~\cite{Gorham:2008dv} which is an air-borne balloon observatory flying over Antarctica. ANITA uses Askaryan radiation from the ice to look for cosmic neutrinos. ANITA has so far  detected two events that resemble the signal of $\nu_\tau$ or $\bar{\nu}_\tau$ coming from the zenith angles of $117.4^\circ\pm 0.3^\circ$~\cite{Gorham:2016zah} and $125^\circ\pm 0.3^\circ$~\cite{Gorham:2018ydl} with energies of $0.6 \pm 0.4$~EeV and $0.56^{+ 0.4}_{-0.2}$~EeV, respectively. At such high energies the neutrino nucleus scattering cross section is relatively large, leading to a mean free path much smaller than the size of the chords corresponding to these zenith angles.

As a result, these two events are considered anomalous and call for an explanation.  Within the standard model, some possible explanations have been provided in terms of coherent transition radiation from the geomagnetically-induced current in cosmic-ray air showers~\cite{deVries:2019gzs}, transition radiation from showers crossing the interface between Earth and air~\cite{Motloch:2016yic} and reflection of radio waves, without phase inversion, off Antarctic sub-surfaces~\cite{Shoemaker:2019xlt} (see~\cite{Anchordoqui:2019utb} for a summary; see also~\cite{Dasgupta:2018dzp,Prohira:2018mmv} for a detailed study of the reflection of spherical waves from curved surfaces including the roughness of surface and possible misidentification of the signals). It is also intriguing to entertain the possibility of finding the footprints of new physics in the two anomalous ANITA events. Various beyond standard model scenarios have been developed in the literature to explain these two events~\cite{Cherry:2018rxj,Anchordoqui:2018ucj,Huang:2018als,Dudas:2018npp,Connolly:2018ewv,Fox:2018syq,Collins:2018jpg,Romero-Wolf:2018zxt,Chauhan:2018lnq,Anchordoqui:2018ssd,Heurtier:2019git,Hooper:2019ytr,Cline:2019snp,Esteban:2019hcm,Heurtier:2019rkz,Chipman:2019vjm,Borah:2019ciw,Abdullah:2019ofw}. A class of these scenarios introduce a new particle which can traverse the Earth with a mean free path larger than that of the standard model neutrinos. The new particle converts to $\nu_\tau$ in the vicinity of the detector giving rise to the signal. The model that we are proposing in this paper belongs to this class of scenarios. We introduce a gauge $L_e-L_\tau$ symmetry with a gauge boson $Z'$ of mass 100-200 GeV with a coupling below the LEP bound. We also introduce a pair of Weyl fermions $N_1$ and $N_2$ with a coupling of the form $Z'_\mu \bar N_1 \gamma^\mu N_2$. The $N_1$ particles can be produced in the energetic sources such as AGN (Active Galactic Nuclei) via interactions with the electrons inside the source. Another possibility  for the $N_1$ flux production is the decay of superheavy dark matter to the $\bar N_1 N_1$ pairs. The $N_1$ particles, being stable, traverse through the cosmos and arrive at the Earth. Thanks to the coupling to the $L_e -L_\tau$ gauge boson, they interact with the electrons inside the Earth converting to $N_2$. Subsequently, $N_2$ decays into $N_1$ and a pair of leptons including $\nu_\tau$ and $\bar\nu_{\tau}$ which account for the observed events.
 
The $N_1$ particles being neutral and stable can be a suitable dark matter candidate. In our scenario, they are produced thermally in the early Universe and their abundance  is set by co-annihilation with $N_2$ via  a freeze-out scenario. The small  splitting between the $N_1$ and $N_2$ masses in our model is natural.
 
The paper is organized as follows. In sect.~\ref{scenario}, we introduce our scenario and determine the range of the relevant parameters that can explain the two anomalous ANITA events while avoiding the bounds from various experiments on the new particles introduced in this scenario. In sect.~\ref{model}, the full gauge invariant model embedding the scenario is described. In sect.~\ref{ILC}, the prospects for testing the model at ILC is described. In sect.~\ref{N1production}, we propose some mechanisms for the production of the initial $N_1$ flux. In sect.~\ref{Zenith}, after a brief discussion of the ANITA events, we discuss the energy and the zenith angle dependence of the emergence of a $\nu_\tau$ and $\bar \nu_\tau$ in the vicinity of the detector within our scenario and then proceed with discussing how the IceCube and Auger bounds as well as the non-observation of events from other zenith angles by ANITA can be explained. The summary and conclusion are presented in sect.~\ref{Sum}.

\section{The model and its predictions for colliders}

In section \ref{scenario}, we first present the scenario explaining the two events observed by ANITA. In section \ref{model}, we  show how this scenario can be embedded within a gauge invariant model. In section \ref{ILC}, we shall discuss the implication of the model for different observations such as the ILC searches and neutrino oscillation. Finally, in section \ref{N1production}, we briefly discuss possible sources for $N_1$ production. The implications for ANITA and IceCube will be further investigated in the next section.
 
\subsection{The scenario\label{scenario}}
 
\begin{figure}[h!]
\centering
\subfloat[]{
\includegraphics[width=0.32\textwidth]{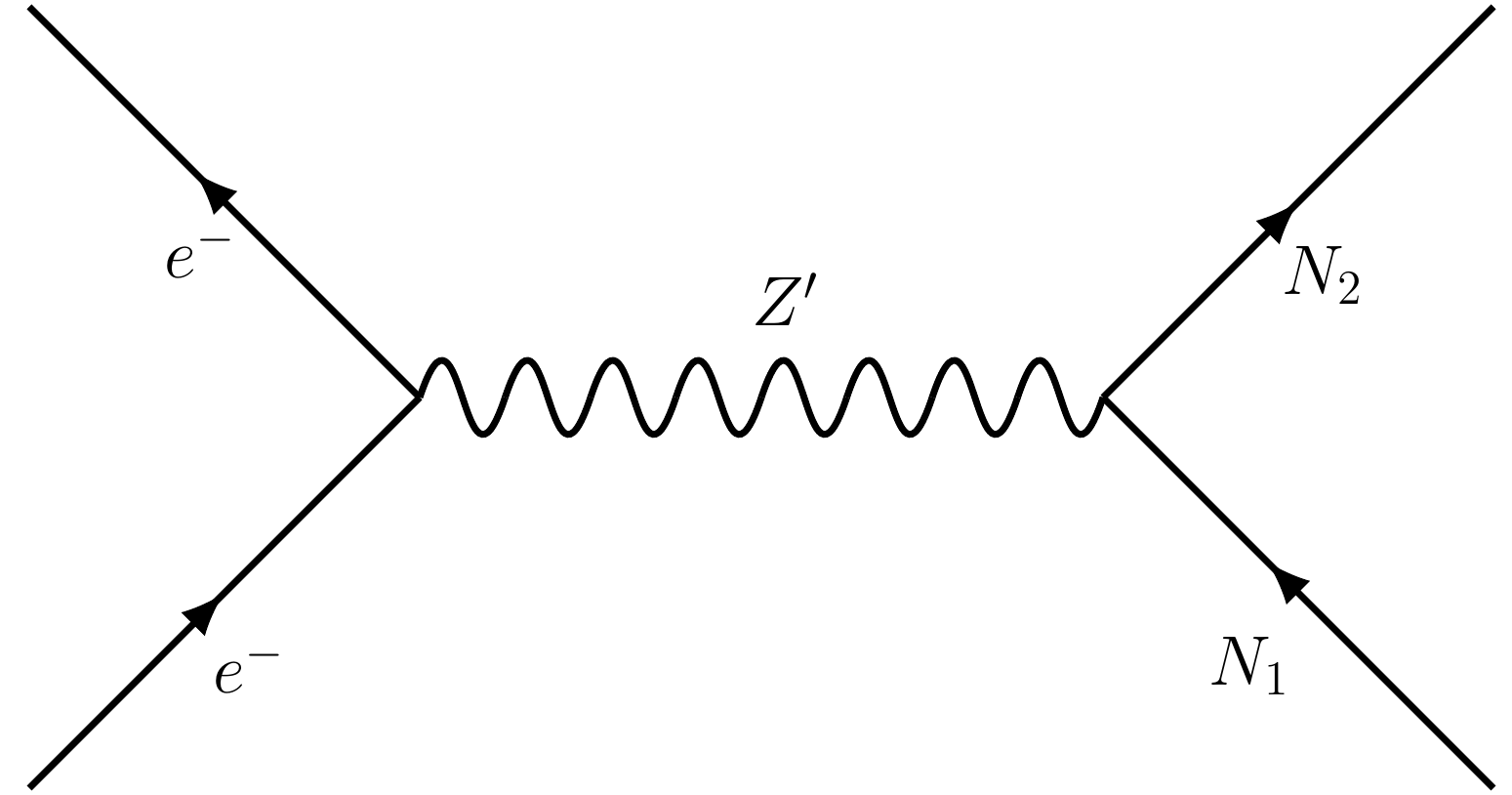}
\label{fig:eN1}
}\hspace{0.5cm}
\subfloat[]{
\includegraphics[width=0.32\textwidth]{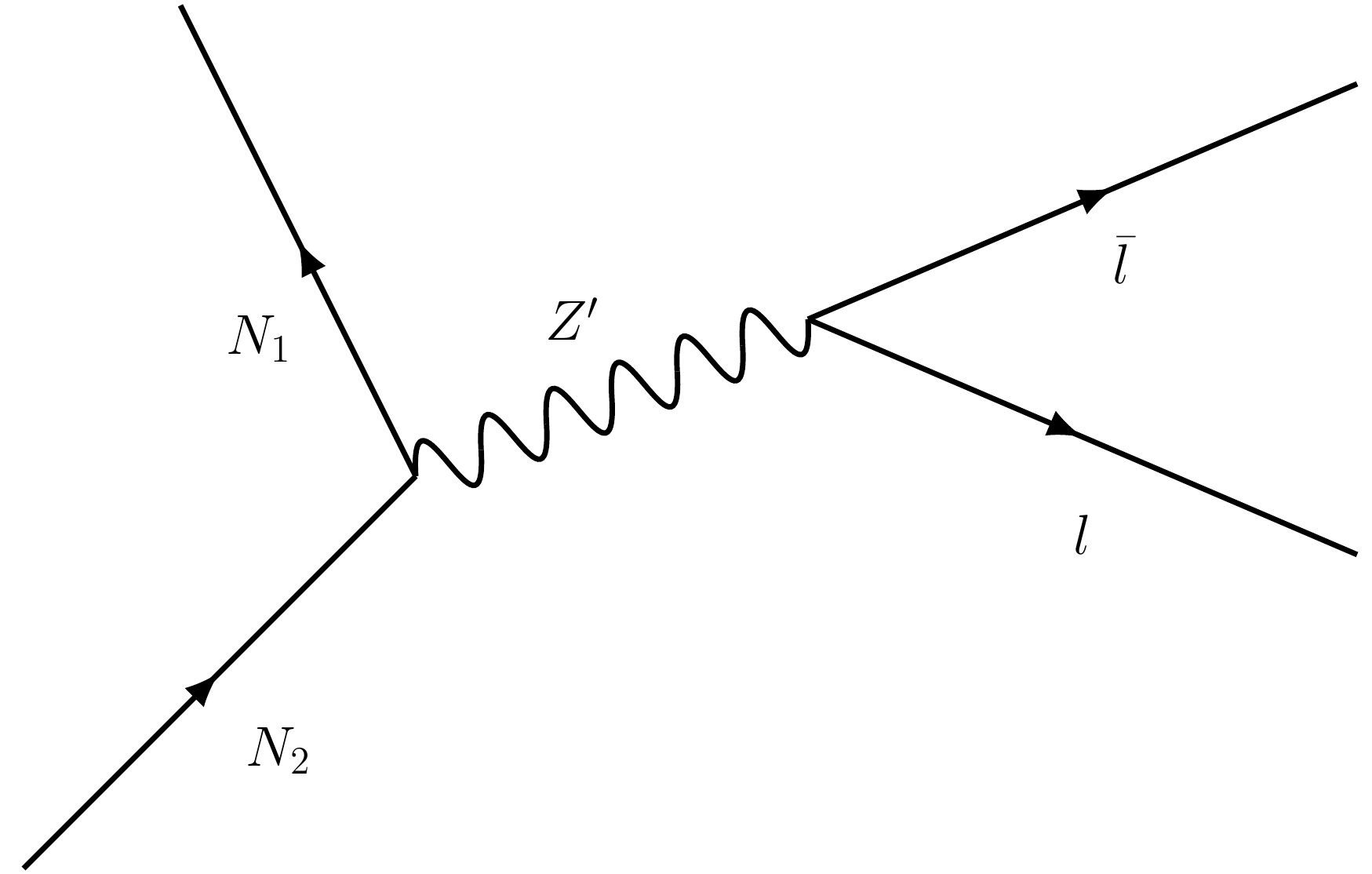}
\label{fig:N2}
}\hspace{0.5cm}
\subfloat[]{
\includegraphics[height=0.27\textwidth]{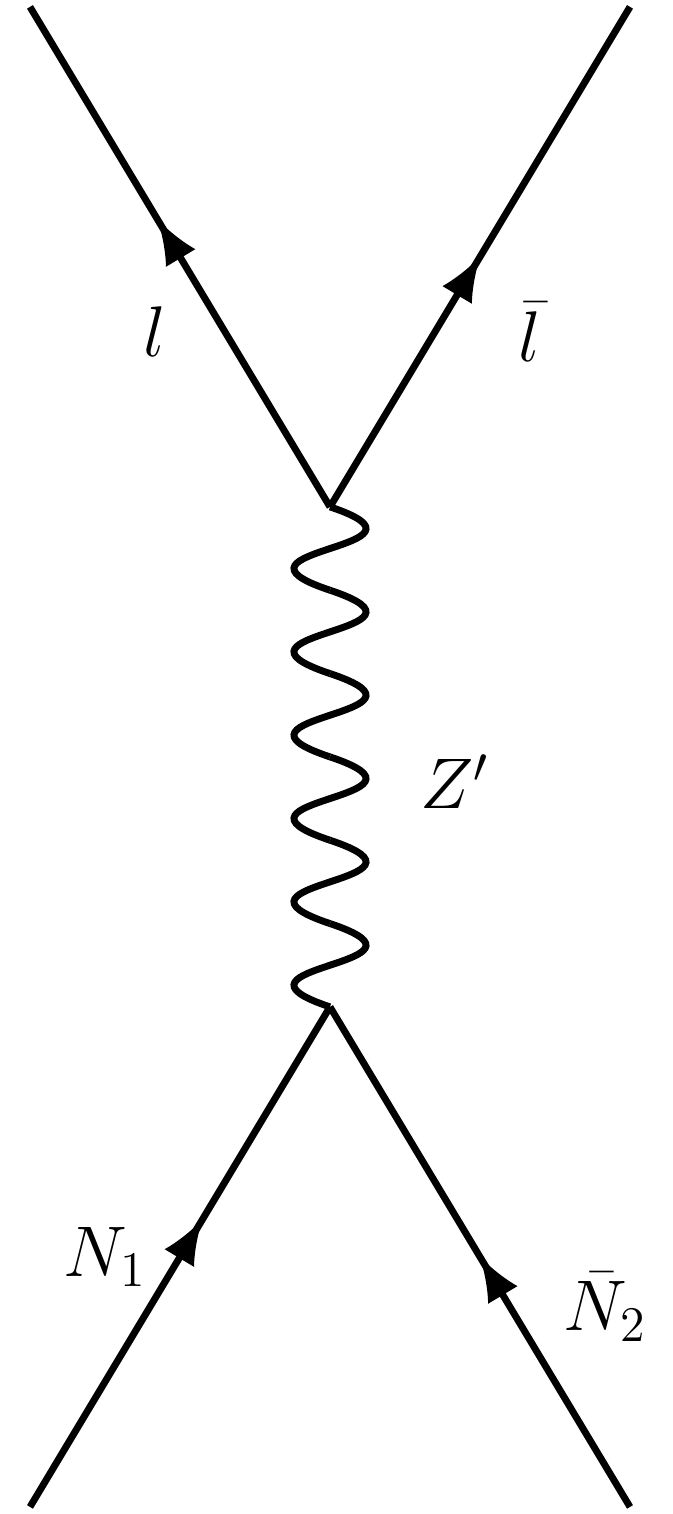}
\label{fig:N1N2}
}
\caption{\label{fig:feyn} Feynman diagrams of a) $e^-N_1$ scattering; b) $N_2$ decay, with $l=e,\nu_e,\tau,\nu_\tau$; c) coannihilation of $N_1\bar{N}_2\to l\bar{l}$ with $l=e,\nu_e,\tau,\nu_\tau$.}
\end{figure}

In our scenario, there are two new Weyl fermions $N_1$ and $N_2$ which couple to a new heavy gauge boson $Z^\prime$ as
\be  g_N {Z}_\mu^\prime \bar{N}_2  \sigma^\mu N_1 +{\rm H.c.}~, \label{coupling}\ee
where $\sigma^\mu=(\mathbb{1},\vec{\sigma})$ is the $2 \times 2$ Pauli matrix four vector. We take $N_2$ (with mass $M_2$) to be heavier than $N_1$ (with mass $M_1$). In order for the scenario to work,  $Z^\prime_\mu$ also  has to couple to $\nu_\tau$ as well as to the matter fields. An elegant way to obtain these gauge couplings is to gauge $L_e-L_\tau$ with a gauge coupling of $g_{e-\tau}$. As a result, a beam of $N_1$ in matter can interact with the electrons via a $t$-channel $Z^\prime$ exchange converting $N_1$ to $N_2$ as shown in Fig. \ref{fig:eN1}. Subsequently, $N_2$ can decay into $N_1$ plus $\nu_\tau \bar{\nu}_\tau$, $\tau \bar{\tau}$, $\nu_e \bar{\nu}_e$  or $e \bar{e}$ as shown in Fig. \ref{fig:N2}: $N_2 \to N_1 \nu_\tau\bar{\nu}_\tau,N_1 \tau \bar{\tau},N_1 e \bar{e},N_1 \nu_e \bar{\nu}_e$. The $Z_2$ parity under which $N_1 \to -N_1$, $N_2 \to -N_2$ (but the rest of the fields are even) makes $N_1$ stable.
 
Similarly to~\cite{Cherry:2018rxj}, we assume that some sources produce a flux of $N_1$ particles. Some examples of such  sources can be the following: \textit{i}) the decay of very heavy dark matter particles: ${\rm DM}\to N_1 \bar{N}_1$; \textit{ii}) the collision of electron electron (or electron proton) in the far away energetic sources such as AGNs or GRBs, producing $Z^\prime$ which decays to the $\bar{N}_1N_2$ and $\bar{N}_2N_1$ pairs: $e^-+e^- (p^+) \to e^-+e^- (X)+Z^\prime \to e^-+e^- (X)\bar{N}_1 N_2, e^-+e^- (X)\bar{N}_2 N_1$ (where $X$ are parton jets). $N_2$ subsequently decays into $N_1$. \textit{iii}) $\nu+{\rm nucleons} \to \nu +Z'+X$ in collapsars (i.e., the progenitors of chocked GRBs). Notice that in the Feynman diagrams of these processes, $Z'$ couples only to the leptonic lines.

The energetic flux of $N_1$ entering the Earth can interact with the electrons via a $t$-channel $Z^\prime$ exchange, as shown in Figure~\ref{fig:eN1}, with a  cross section of
\be \sigma(N_1+e \to N_2+e) =\frac{g_N^2 g_{e-\tau}^2}{8 \pi}s \int_{-1}^{1}\frac{4+(1+\cos \theta)^2}{(s(1-\cos \theta)/2+m_{Z'}^2)^2}\,d \cos \theta \label{sigma}~,
\ee
where $s$ is a Mandelstam variable: $s=2 m_e E_{N_1}={\rm TeV}^2~(E_{N_1}/{\rm EeV}).$ Notice that in Eq. (\ref{sigma}) we have neglected the masses of $N_1$ and $N_2$ in comparison to their energy. At this approximation, $ \sigma(N_1+e \to N_2+e)\simeq  \sigma(N_2+e \to N_1+e)$. Notice that for $s\gg m_{Z'}^2$ at $1-2m_{Z'}^2/s<\cos \theta<1$, the integrand in Eq (\ref{sigma}) is enhanced so $\sigma \simeq (2g_N^2g_{e-\tau}^2/ \pi m_{Z'}^2)$ becomes independent of $s$. The scattering converts $N_1$ to $N_2$ which can subsequently decay into $N_1$ with a total decay rate at the lab frame (see Figure~\ref{fig:N2}) given by 
\be \Gamma_{tot}=\frac{g_N^2 g_{e-\tau}^2}{10\pi^3} \frac{(M_{2}-M_1)^5}{ m_{Z'}^4} \left( \frac{M_{2}}{E_{N_2}}\right).\label{GamTOT} \ee 
The ratio in the last parenthesis is the inverse of the boost factor taking care of time dilation. The decay modes $N_2\to N_1 \nu_\tau \bar{\nu}_\tau$ and $N_2\to N_1 \tau \bar{\tau}$ can contribute to the ANITA events. That is about a half of decaying $N_2$ produce $\nu_\tau$: $B={\rm Br}(N_2\to N_1 \nu_\tau \bar{\nu}_\tau) +Br(N_2\to N_1 \tau \bar{\tau})$ which equals 1/2 for $m_\tau \ll M_2-M_1$. In the case that $2m_\tau >M_2-M_1$, $B=1/4$. The probability that an $N_1$ entering the Earth from an angle corresponding to a chord of length $L$ produces a $\nu_\tau$ or $\bar\nu_\tau$ in the vicinity of ANITA ({\it i.e.,} within the mean free path of $\nu_\tau$, $\tau_\nu =(\sigma_{SM}~\rho/m_p)^{-1}$) is given by
\begin{equation}\label{eq:prob}
 P={\rm Min}[1, \Gamma_{tot}\tau_\nu](2B)\int_0^L e^{-\int_0^x \gamma(y) dy}e^{-\int_x^L (\gamma(z) +\Gamma_{tot}) dz} \gamma dx~,
\end{equation}
where $\gamma$ is the inverse of the mean free path of $N_1$ (as well as that of $N_2$): $\gamma(x) =n_e(x) \sigma$ in which $n_e$ is the electron number density of Earth's matter. The factor of $e^{-\int_0^x \gamma(y) dy}$ in the integrand is the probability of $N_1$ to survive scattering up to $x$. $\gamma dx$ is the probability of scattering and converting of $N_1$ into $N_2$ in the element $dx$.  $e^{-\int_x^L (\gamma(z) +\Gamma_{tot}) dz}$ is the probability that the produced $N_2$ does not scatter or decay up to reaching the surface. Considering that each $N_2\to N_1 \nu_\tau \bar{\nu}_\tau$ or $N_2\to N_1 \tau \bar\tau$ produces a pair of $\nu_\tau$ and $\bar{\nu}_\tau$, $2B$ is the average number of $\nu_\tau$ and $\bar{\nu}_\tau$ produced at each $N_2$ decay. Finally ${\rm Min}[1, \Gamma_{tot}\tau_\nu]$ is the probability that $N_2$ decays in the vicinity of ANITA. Assuming constant density along the propagation chord, we obtain
\begin{equation}
P \simeq (2B){\rm Min}[1, \Gamma_{tot}\tau_\nu] \frac{\gamma}{\Gamma_{tot}}\left[e^{-L\gamma}-e^{-L(\gamma+\Gamma_{tot})}\right]~. 
\end{equation}
$P$ is maximal for $\Gamma_{tot}\sim 1/\tau_\nu \sim (500~{\rm km})^{-1}$ and $\gamma \sim 1/L \sim (5000~{\rm km})^{-1}$. With these values, $P$ will be  a few percent. Let us check whether in our scenario these values can be obtained. 
  
The gauge boson of the $L_e-L_\tau$ symmetry can be relatively light. Since this gauge boson is not coupled to the quarks at the tree level, from the LHC no significant bound can be set on $m_{Z'}$. The bound from LEP~\cite{Freitas:2014jla} is
\begin{eqnarray} \label{bAnds} \frac{g_{e-\tau}}{m_{Z'}} <  2.0 \times 10^{-4} {\rm GeV}^{-1} ~~~  & {\rm for } ~~~~ & 200~{\rm GeV}<m_{Z'} \cr
  	\ \frac{g_{e-\tau}}{m_{Z'}} <  6.9\times 10^{-4} {\rm GeV}^{-1} ~~~  & {\rm for } ~~~~ & 100~{\rm GeV}<m_{Z'}<200~{\rm GeV} .\end{eqnarray}
Taking $m_{Z'}\sim 100-200$~GeV and saturating the bound on $g_{e-\tau}$ [{\it i.e.,} $g_{e-\tau}=6.9\times 10^{-2}(m_{Z'}/ 100~{\rm GeV})$] independently of the value of $m_{Z'}$, we find
$$\sigma(e+N_1 \to e+N_2)= \sigma(e+N_2 \to e+N_1)= 10^{-34}g_N^2~{\rm cm}^2~,$$
which for the Earth mantle with $\rho=4~{\rm gr/cm}^3$ with almost equal proton and neutron composition ({\it i.e.,} with $n_e=\rho/(2 m_p)$), the mean free path will be $\gamma^{-1}=(n_e \sigma)^{-1}=8 \times 10^4~{\rm km}/g_N^2$. Taking $g_N\sim 3$, $\gamma^{-1}$ will be close to the chord size. With $g_{e-\tau}/m_{Z'}=6.9\times 10^{-4}~{\rm GeV}^{-1}$ and $g_N=3$, $\Gamma_{tot}\sim (500~{\rm km})^{-1}$ implies $M_2 -M_1 \sim 0.7 ~{\rm GeV}(10~{\rm GeV}/M_2)^{1/5}$. With such small splitting in the early universe, $N_1 \bar N_2$ as well as $N_2 \bar N_1$ pairs can coannihilate, via a $s$-channel interaction as shown in Figure~\ref{fig:N1N2}, with a cross section
\be \left\langle \sigma (N_1 \bar N_2 \to {\rm lepton~pairs})v\right\rangle \sim 3\, \frac{g_N^2 g_{e-\tau}^2M_1^2}{\pi m_{Z'}^4}=1.6 \times 10^{-35}\left(\frac{M_1/m_{Z'}}{0.1}\right)^2 {\rm cm}^2\, \gtrsim 1 ~{\rm pb}~,\ee 
so $N_1$ particles will not overclose the universe and their contribution to dark matter can be $\mathcal{O}(1)$.

The energy of $N_2$ will be about half of the energy of $N_1$ as the other half would be carried away by the electron on which  scattering has taken place. The energies of $\nu_\tau$ and $\bar \nu_\tau$ produced at $N_2 \to N_1 \nu_\tau \bar{\nu}_\tau$ will be  $\mathcal{O}[(E_{N_2}/2)(M_2-M_1)/M_1]$. Taking $(M_2-M_1)/M_1 \sim 0.1$, the energy of the final neutrinos will be about $\sim 2-3 \%$ of the energy of the initial $N_1$. The energies of $\nu_\tau$ and $\bar{\nu}_\tau$ from the $\tau$ decay after $N_2 \to N_1 \tau \bar{\tau}$ will be further suppressed. For $M_2-M_1 \sim M_2$, the energy of the final $\nu_e$ could be $E_{N_1}/6$ but we should have then devised another annihilation mode for $N_1 \bar N_1$ to prevent the overclosure of the universe. Moreover as we shall see in the next subsection, obtaining quasi-degeneracy is more natural than obtaining $M_2-M_1 \sim M_2$ from the model building point of view.

Let us summarize the main features of our scenario. Stable $N_1$ particles with a mass of $\sim 10$ GeV come to Earth from cosmic sources. These $N_1$ particles scatter on the electrons in the Earth via a $t$-channel exchange of a $Z'$ gauge boson with a mass of $\sim 100$ GeV, converting $N_1$ to $N_2$ with a mass splitting $M_2-M_1 \sim$ GeV. $N_2$ decays back to $N_1$ plus a lepton pair via the same interaction. These pairs can be observed by ANITA. The gauge interaction that we have taken  is $L_e-L_\tau$.\footnote{In principle, instead of this combination, we could gauge $B-3L_\tau$ (or some other anomaly free combination of $B$, $L_\tau$, $L_\mu$ and $L_e$). In that case, the bounds from the LHC on the new $Z'$ should be considered. The bounds reported by CMS and ATLAS are for sequential $Z'$ with a coupling similar to the SM $Z$. With gauge coupling as large as 0.1, the lower bound on $m_{Z'}$ from \cite{Zucchetta:2019afp} implies the $N_1$ proton scattering cross section will be too small. However, regions of the parameter space with $m_{Z'}\sim 100$ GeV and couplings giving rise to sufficiently small mean free path for $N_1$ might still be allowed.}

\subsection{The model \label{model}}

Let us now embed the scenario within a UV complete model. The $L_e -L_\tau$ combination that we have chosen to gauge is  an anomaly free combination. There is a vast literature on $L_e -L_\tau$ gauge models with a massive $Z'$ \cite{gauges}. We shall not repeat this part. In the following, we will instead build a model that gives rise to the coupling in Eq.~(\ref{coupling}). Let us define $\psi_1=(N_1+N_2)/\sqrt{2}$ and $\psi_2=(N_1-N_2)/\sqrt{2}$. Under the new $U(1)$, we assign opposite charges to these Weyl fermions. As a result, the $U(1)^3$ anomaly cancels. Moreover, the gauge interaction takes the desired form:
$$ g_N(\bar{\psi}_1 \sigma^\mu \psi_1-\bar\psi_2 \sigma^\mu \psi_2)Z_\mu^\prime= g_N (\bar{N}_1 \sigma^\mu N_2 +\bar{N}_2 \sigma^\mu N_1) Z_\mu^\prime~.$$ 
To reproduce our scenario, we should make sure that $N_1$ and $N_2$ are mass eigenstates with a small mass splitting. Let us define $c$ to be a $2\times 2$ asymmetric matrix with off-diagonal elements equal to $\pm 1$. A mass term of 
$$m(\psi_1^T c\psi_2+\psi_2^Tc\psi_1)=m(N_1^TcN_1-N_2^TcN_2)~,$$ preserves the gauge symmetry. To create a splitting between $M_1$ and $M_2$ we need a mass term of $\psi_1^T c\psi_1+\psi_2^T c\psi_2$ which breaks the gauge symmetry. This can be achieved by introducing a complex $\Phi$ with a charge under new gauge symmetry equal to twice that of $\psi_2$. We can then write Yukawa couplings $Y_1 \Phi \psi_1^Tc \psi_1$ and $Y_2 \Phi^* \psi_2^Tc \psi_2$. For general $Y_1$ and $Y_2$, the mass terms of form $N_1^TcN_2$ appear deviating $N_1$ and $N_2$ from mass eigenstates. Imposing a symmetry under which $\psi_1 \leftrightarrow \psi_2$ and $\Phi \leftrightarrow \Phi^*$ (and $Z' \leftrightarrow - Z'$ and $e \leftrightarrow \tau$) leads to $Y\equiv Y_1=Y_2$ which in turn results in mass terms $Y\langle \Phi\rangle (N_1^TcN_1+N_2^TcN_2)$. Thus, 
$$M_1=|m+Y\langle \Phi\rangle| \ \ \  {\rm and}    \ \ \ M_2=|m-Y\langle \Phi\rangle|~.$$
Taking $m\sim 10$~GeV and $Y\langle \Phi\rangle\sim 0.5$~GeV (or the other way around), the quasi-degeneracy of $N_1$ and $N_2$ can be naturally explained. The $\Phi$  particle can be produced in the early universe. Taking the $\Phi$ particles to be heavier than $2M_2$, they can decay fast to $\bar N_2N_2$ and $\bar N_1 N_1$ avoiding the bounds from cosmology on light degrees of freedom.

The new scalar $\Phi$ can have a coupling of form $\lambda_{H \Phi}|H|^2 |\Phi|^2$ with the SM Higgs. This will induce a mixing between the Higgs and $\Phi$. Moreover for $2m_\Phi<m_H$, it can lead to a new decay mode for the SM Higgs: $H\to \Phi\bar{\Phi}\to N_1 \bar N_1 N_2 \bar{N}_2 \to 2N_1 2\bar N_1 l\bar l l' \bar l'$. Taking $\lambda_{H \Phi}\ll m_\tau /\langle H \rangle$, this decay mode can be neglected. At the two loop level, this coupling induces a tiny $N_2^T cN_1$ mass term suppressed by $\lambda_{H \Phi}Y g_N g_{e-\tau}(m_\tau^2-m_e^2)/m_{Z'}^2$ which means that the ``real" mass eigenstates
$\tilde{N}_1$ and $\tilde{N}_2$ will slightly deviate from $N_1$ and $N_2$ creating $\bar N_1 \sigma^\mu N_1 Z_\mu'$, $\bar N_2 \sigma^\mu N_2 Z_\mu'$ as well as $ N_1^Tc N_2\Phi$ with suppressed couplings with no dramatic consequence for our  scenario.

At one loop level, a kinetic mixing between $Z'$ and the photon will be created with a mixing of $\mathcal{O}(g_{e-\tau}e/(16 \pi^2))$. This will lead to the $N_1$ scattering off nucleons with a cross section of $\sim [g_N g_{e-\tau} e^2/(16 \pi^2 g'^2)]^2 \sigma(\nu+{\rm nucleon})\sim 10^{-37}{\rm cm}^2$ at the energy of $\mathcal{O}(1)$~EeV. The corresponding mean free path will be too large to be relevant.

Notice that the scattering of non-relativistic $N_1$ off the nuclei (as is the case assuming dark matter role for $N_1$ particles) is kinematically forbidden: in this case $E_{N_1}\sim M_{1}(1+v^2/2)$, where $v\sim10^{-3}$ is the  average velocity of dark matter particles in the solar system relative to the Earth.  Thus, $N_1 +( e^-~\textrm{or nucleus}) \to N_2 +( e^-~\textrm{or nucleus})$ is not kinematically allowed and  the bounds from the direct dark matter searches can be therefore avoided.

\subsection{The predictions for the ILC and neutrino oscillation experiments \label{ILC}}

Let us now discuss the implications of this model for various experiments other than neutrino telescopes. As shown in~\cite{Freitas:2014jla}, $Z'$ with values of $g_{e-\tau}/m_{Z'}$ of interest to us can be easily discovered by ILC via the process $e^-e^+ \to \gamma Z'$ and subsequently $Z' \to e^-e^+$. In our model $g_N\gg g_{e-\tau}$ so $Z'$ will mainly decay into $N_1\bar N_2$ and $N_2\bar N_1$. The four-momenta of the initial $e^-$ and $e^+$ are known and the four-momenta of the photon in $e^-e^+ \to \gamma Z'$ can be measured. Thus, independent of the final decay products of $Z'$, we expect a peak at $(P_{e^-}+P_{e^+}-P_\gamma)^2= m_{Z'}^2$. As a result, by measuring the four momenta of $\gamma$, the mass of $Z'$ can be reconstructed. Moreover, the height of the peak gives $g_{e-\tau}^2$.

$N_1$ produced at the $Z'$ decay will appear as missing energy but $N_2$ will decay with a decay length of $1~{\rm cm}~(E_{N_2}/20~{\rm GeV})$.
\begin{itemize} 
\item  For $M_2-M_1\gg 2m_\tau$  (for $M_2-M_1< 2m_\tau$) in about  $1/3$  ($1/2$) of cases, we expect $N_2$ decays into $N_1 \nu_e \bar{\nu}_e$ or into $N_1 \nu_\tau \bar{\nu}_\tau$ which again appear as missing energy so the signature at ILC will be a $\gamma$ with $(P_{e^-}+P_{e^+}-P_\gamma)^2=m_{Z'}^2$ plus missing energy. 

\item For $M_2-M_1\gg 2m_\tau$  (for $M_2-M_1< 2m_\tau$) in about $1/3$  ($1/2$) of cases, $N_2$ decays into $N_1 e^-e^+$ so the signature will be a photon as described above and an $e^-e^+$ pair at a displaced vertex. The displacement is given by $N_2$ decay length which gives information on $(M_2-M_1)^5 g_N^2$. The energy of $Z'$ will be $(s+m_{Z'}^2)/(2 \sqrt{s})$ and the energy of $e^-$ or $e^+$ will be around $[(s+m_{Z'}^2)/(8 \sqrt{s})](1-M_1/M_2)$. Taking $\sqrt{s}=500$~GeV and $1-M_1/M_2\sim 0.1$, the energies of $e^-$ and $e^+$ will be $\sim 6$~GeV. If the detector can register such low energy electron and positron and measure their energy momentum, independent information on $M_1$ and $M_2$ can be extracted. If $(M_2-M_1)/M_2 \sim 1$ (which means $m\sim Y\langle \Phi \rangle$), the energies of $e^-$ and $e^+$ can be much larger and their detection will be guaranteed.

\item  For $M_2-M_1\gg 2m_\tau$   in about $1/3$ of cases, the decay leads to $\tau^-\tau^+$ pair, $e^-e^+ \to \gamma Z' \to \gamma N_1\bar N_1 \tau^-\tau^+$. Similar consideration applies to this mode, too.
\end{itemize}

Another potential experiment where the effects of $Z'$ can show up is the $(g-2)_e$ measurements. The contribution to $(g-2)_e$ will be of order of $(g_{e-\tau}^2/16 \pi^2)(m_e^2/m_{Z'}^2) \sim 2.5 \times 10^{-15}$ which is well below the sensitivity of the current experiments~\cite{g-2}.
 
For neutrino oscillation and low energy neutrino scattering experiments, the effects of the new gauge coupling can be described by the following effective four Fermi interaction
\be 2\sqrt{2} \epsilon^e G_F (\bar{\nu}_e \gamma^\mu P_L\nu_e -\bar{\nu}_\tau \gamma^\mu P_L\nu_\tau) (\bar e \gamma_\mu e)~,\ee 
where for $g_{e-\tau}/m_{Z'}$ saturating the bound in Eq.~(\ref{bAnds}), $\epsilon^e \simeq 0.01$. The current bounds do not rule out such tiny values of $\epsilon^e$~\cite{Farzan:2017xzy} but improvements by a factor of 5 in the solar neutrino electron scattering measurements by experiments such as BOREXINO can test this value of $\epsilon^e$~\cite{Agarwalla:2019smc} (see also, \cite{Heeck:2018nzc}). (Super)PINGU may also probe such small $\epsilon^e$~\cite{Choubey:2014iia}.

At the loop level, the quark-$Z'$ coupling will be of the order of $e^2 g_{e-\tau}/(16 \pi^2)$. Considering the strong bounds from LEP on $g_{e-\tau}$, the values of loop-level coupling of $Z'$ to quarks will be even below the reach of HL-LHC \cite{Han}.

\subsection{Possible production mechanisms for $N_1$ \label{N1production}}

Let us now speculate about the possible sources of the $N_1$ flux. In the following, we briefly discuss three examples:
 
\begin{itemize}

\item As discussed before $N_1$ can contribute to the dark matter but it does not need to constitute the whole dark matter. Let us suppose the main component of dark matter is superheavy particles. The high energy flux of $N_1$ can be produced from the decay of a superheavy dark matter particle. The interaction between $N_1$ and the dark matter should respect the new $U(1)$ gauge symmetry. An economic solution is to take the dark matter to be a scalar $(\Phi)$ singlet under the gauge symmetry with a coupling of form
$$\Phi (\psi_1^T c\psi_2 + \psi_2^T c\psi_1)=\Phi (N_1^TcN_1-N_2^TcN_2)~.$$
This implies that the $\Phi$ decay will, along the $N_1$ flux, produce a $N_2$ flux with exactly the same intensity and spectrum. Since the dark matter is non-relativistic, the energies of $N_1$ and $N_2$ will be monochromatic and both equal to $m_\Phi/2$. Notice that this flux will be non-transient; moreover, we expect a higher flux from regions such as the galactic center where the dark matter concentration is higher.

\item $N_1$ produced in the colliding relativistic jets in sources such as AGNs and collapsars. Scattering of high energetic $e^-$ on $e^-$ or $p^+$ can produce $Z'$: $e^- e^- (p^+)\to e^-e^- (X)Z^\prime$ and subsequently, $Z^\prime \to \bar{N}_1 N_2, \bar{N}_2 N_1$. Here, we again expect equal fluxes for $N_2$ and $N_1$ but their energy spectrum will be continuous.  Notice that in such a powerful source processes $e^-e^-(p^+)\to e^-e^-(X)\gamma,Z$ (where $X$ are parton jets) can also take place which can in principle lead to additional neutrino as well as photon signal. As we discussed before, in order to have $\nu_\tau$ with an energy of $\sim 0.6$~EeV, the energy of the initial $N_1$ should be $\sim (20-30)$ EeV which means that the colliding charged particles in the jets must have energies above $\mathcal{O}(50)$~EeV. This in turn implies that if the accelerated protons are leaked out of the source, they will be stopped by the intermediate CMB photons and cannot reach us. The source may be too dense to allow the high energy protons to leak out (especially for the collapsars) but the $N_1$ particles can easily come out and reach our detectors. Studying the relevant  bounds and possible mechanisms  to circumvent those bounds  (such as absorption at source) are beyond the scope of the present paper.
\item As is well-known, the collapsing very heavy stars with a mass above $40M_\odot$ can be considered as a transient source of very high energy $\nu_e$ and $\nu_\mu$ and their antiparticles. For energies higher than EeV, the envelope of the star (just like the earth) will be opaque to neutrinos. In our model, $\nu_e$ with energies above 10 EeV can scatter off nuclei (by the $t$-channel electroweak  $Z$  boson exchange) and emit $Z'$ with energies of few EeV with a cross section 
$$\sigma(\nu_e+{\rm nucleus} \to \nu_e +X+Z') \sim \frac{g_{e-\tau}^2}{16 \pi^2}\, \sigma(\nu_e+{\rm nucleus} \to \nu_e +X)~.$$
The decay of  $Z'$ can produce the needed transient $N_1$ flux. The majority of $N_1$ can come out of the envelope without hinderance.

\end{itemize}
In both cases, $N_2$ will decay to $N_1$ and a pair of leptons. Since the mass splitting of $N_2$ and $N_1$ is small, the spectrum of the secondary $N_1$ will be similar to that of $N_2$ so in practice the flux of $N_1$ will be doubled. The energy of the produced $\nu_e \bar{\nu}_e$ and $\nu_\tau \bar{\nu}_\tau$ will be smaller by a factor of $(M_2-M_1)/(2M_2)$. $\nu_e$ and $\nu_\tau$ will oscillate producing $\nu_\mu$, too.

\section{Consequences of the scenario for ANITA observations\label{Zenith}} 

In this section after a short description of the anomalous events observed by ANITA, we quantify the characteristics (the energy and angular distributions) of the flux of $N_1$ particles that can give rise to the two anomalous events observed by ANITA and at the same time avoiding the bounds from IceCube and Auger as well as from ANITA itself. We then comment on the observation of more events in the future.

\subsection{The ANITA experiment and the anomalous events\label{Anita}} 

The ANITA experiment is a balloon radio wave detector at Antarctic~\cite{Gorham:2008dv}. Although primarily the experiment was designed to detect the radio wave emission from Askaryan effect in ice due to high energy neutrino interaction~\cite{Gorham:2006fy}, ANITA's detectors can be triggered also by the impulsive radio wave emission from the dipole radiation of the charge asymmetry developing in the showers originated by the either down-going or horizontally propagating ultra high energy cosmic rays~\cite{Hoover:2010qt} (the Askaryan effect radiation from the shower is quite small in comparison to the dipole radiation). The charge asymmetry develops in the shower due to the geomagnetic field which at the South Pole is predominately vertical, leading to a lateral charge separation which radiates radio waves with horizontal polarization. Due to the geomagnetic field configuration and the geometry of the detectors, ANITA is sensitive to showers with zenith angle $\theta_z\gtrsim60^\circ$. However, ANITA is still sensitive to the down-going showers by looking at the reflected radio waves from the ice, although the polarity of the radio signal will be inverted in this case. The man-made (anthropogenic) radio signals mimicking the same polarization as the signal (being either the cosmic rays or the neutrinos) can be rejected due to the small, but non-negligible, horizontal geomagnetic field components at Antarctic. The small horizontal geomagnetic field components produce vertically polarized radio waves such that the proportion of vertical and horizontal polarizations, the direction of the shower propagation and the location of observation have correlation dictated by the precisely known geomagnetic field at the location. Since the ANITA possesses detectors sensitive to both the vertical and horizontal polarizations, the correlation with the geomagnetic field robustly can be measured which provides a powerful handle on tagging the cosmic ray events.

Using the polarity method succinctly described above, ANITA started  data taking looking for neutrinos at very high energies (motivated by and in search for the putative cosmogenic neutrino flux) in three flight periods. In two datasets (the flight periods I and III) anomalous events have been detected. The \textit{anomalous} ANITA events consist of two Earth-emerging showers with almost equal energies: \textit{i}) the event \#3985267 observed during the ANITA-I flight with a  zenith angle of $117.4^\circ$ and an energy of $0.6\pm0.4$~EeV~\cite{Gorham:2016zah}; \textit{ii}) the event \#15717147 registered during the ANITA-III flight from a zenith angle of $125^\circ$ and an energy of $0.56^{+0.3}_{-0.2}$~EeV~\cite{Gorham:2018ydl}. The ANITA-I (ANITA-III) event has been tagged as anomalous among a set of 16 (20) ultra high energy cosmic ray \textit{normal} events expected from down-going cosmic ray showers and on a background of anthropogenic sources and the showers mostly from horizon at the ANITA's balloon location. The probability of the anomalous events being anthropogenic in ANITA-I and ANITA-III are $\simeq4\times10^{-4}$ and $\simeq1.2\times10^{-3}$ (or $\simeq0.015$ in a conservative analysis), respectively. The two anomalous events, not showing any inverted polarization, are consistent with Earth-emerging showers which should arise from the propagation of neutrinos ($\nu_\tau$ or $\bar{\nu}_\tau$) of $\sim$~EeV energy inside the Earth and their subsequent charged-current interaction close to the surface which initiates a shower. Comparing the mean free path of EeV neutrinos in the mantle, $\sim$~500 km, with the propagation chord lengths for the observed zenith angles, $\sim$~5800~km and 7300~km respectively for ANITA-I and ANITA-III events, the traversing probability of neutrinos is $\sim 10^{-9}$. Considering the $\tau$-regeneration inside the Earth, the probability increases to $\sim 10^{-7}$~\cite{Fox:2018syq}. The estimated exposure of ANITA is $2.7~{\rm km}^{2}~{\rm yr}~{\rm sr}$~\cite{Schoorlemmer:2015afa,Fox:2018syq}. Thus, in order to explain the two anomalous ANITA events, within the standard model of particles, an isotropic and diffuse flux of $\nu$ on the Earth $\sim 10^{7}~{\rm km}^{-2}~{\rm yr}^{-1}~{\rm sr}^{-1}$ at $\sim$~EeV is required which is six orders of magnitude larger than the upper limit set by Auger~\cite{Aab:2015kma} and IceCube~\cite{Aartsen:2018vtx} experiments. The assumption of transient neutrino sources also cannot explain ANITA events within the standard model of particles. A transient source with a power-law spectrum extending to lower energies $\sim$~TeV-PeV lead to several events in IceCube, in temporal and spacial correlation with the ANITA events, where no significant excess over the background has been observed~\cite{Pizzuto:2019vyj}. Even a transient and monochromatic flux of neutrinos at $\sim$~EeV will produce lower energy events in IceCube coming from the secondary flux generated when the $\nu_\tau$ component of the flux traverses the Earth~\cite{Pizzuto:2019vyj}. This conflict calls for some Beyond Standard Model (BSM) mechanism allowing for larger traversing probability, either by converting the incident neutrinos to some sterile particle(s) which can propagate across the Earth without absorption, or by assuming the production of neutrinos from a flux of BSM particles incident on Earth which will produce the neutrinos by interacting with matter deeper in Earth (the model proposed in this paper is in fact a combination of these two). However, the BSM scenarios still face a few challenges. Justifying the lack of ANITA events with larger and smaller zenith angles and energies is one of the challenges. The lack of events in Auger and IceCube experiments poses another challenge. Some resolutions for these challenges have already been discussed in the literature, see~\cite{Anchordoqui:2019utb} for a summary. In the following, we describe how these challenges can be addressed in our scenario. 
 
\subsection{Zenith and energy distributions of the expected events\label{dis}} 
       
\begin{figure}[t!]
\centering
\includegraphics[width=0.7\textwidth]{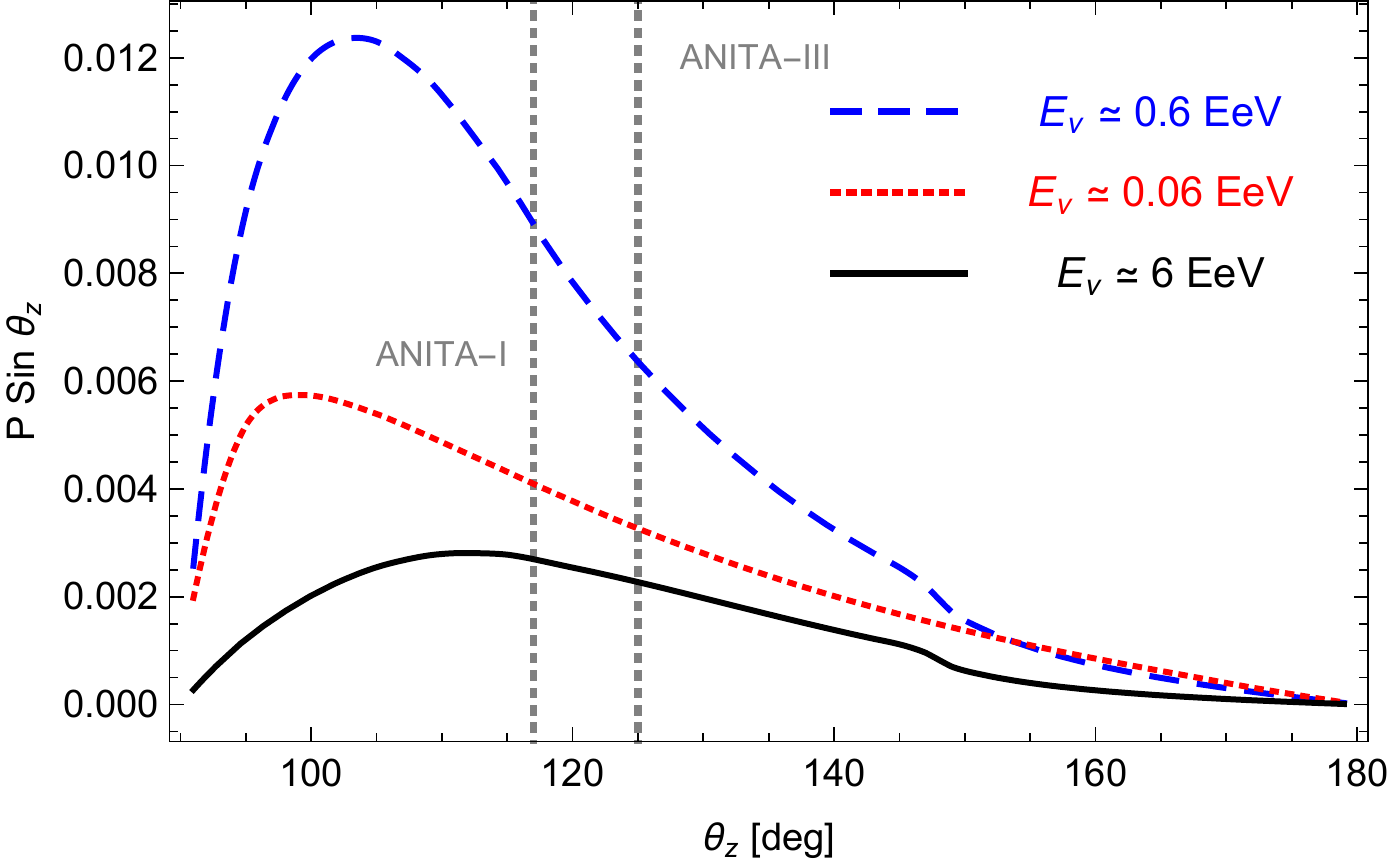}
\caption{\label{fig:prob-z} The success rate $P$ as function of zenith angle $\theta_z$. The success rate has been multiplied by $\sin\theta_z$ which takes into account the solid angle effect. The vertical gray dashed lines show the zenith angles of the observed events by ANITA. The three curves correspond to monochromatic $N_1$ fluxes at the Earth with energies $E_{N_1}=(3.2,32,320)$~EeV, which taking $E_\nu\sim E_{N_1}(M_2-M_1)/(4M_1)$ respectively produce emergent neutrinos with energies $E_\nu\simeq(0.06,0.6,6)$~EeV. For all the curves, the values of the parameters of the model are taken as indicated for the benchmark (A) in Table~\ref{tab:bench}.}
\end{figure}

As discussed in Eq.~(\ref{eq:prob}), the expected zenith angle distribution of the Earth emerging events in our scenario depends on two competing factors: the probability of $N_2$ creation via the scattering of $N_1$ on the Earth's matter ($N_1+e^-\to N_2+e^-$) and the survival probability (or lifetime) of the $N_2$ particles in the propagation from the creation point to the proximity of Earth's surface. Notice that in our scenario the observed neutrinos by ANITA are produced directly from the decay of $N_2$ which produces $\nu_\tau$ or $\tau$, and so the decay should occur within one or two mean free path lengths of neutrinos at $\sim$~EeV. \footnote{Another possibility discussed in~\cite{Fox:2018syq}, not been utilized in our scenario, is the creation of $\nu_\tau$ or $\tau$ with the energy $\gtrsim$~EeV within a few mean free path from the surface such that the $\tau$-regeneration in the Earth's matter produces neutrinos with degraded energy $\sim$~EeV.} The success rate of emerging showers (either $\nu_\tau$ or $\bar{\nu}_\tau$ neutrinos from the $N_2$ decay inside the Earth or the $\tau^\pm$ from the $N_2$ decay close to the surface) can be calculated from Eq.~(\ref{eq:prob}) using the PREM of the Earth~\cite{Dziewonski:1981xy}, for the electron number density at any zenith angle and energy. Figure~\ref{fig:prob-z} shows the success rate $P$ (multiplied by $\sin\theta_z$ to take into account the solid angle effect) as a function of zenith angle $\theta_z$ for a flux of $N_1$ particles with the incident energy such that the emerging neutrino energy is $E_\nu\simeq(0.06,0.6,6)$~EeV, respectively for red-dotted, blue-dashed and black-solid curves. The assumed values of model's parameters for all the curves in Figure~\ref{fig:prob-z} are shown as benchmark (A) in Table~\ref{tab:bench}. In Figure~\ref{fig:prob-z} we have assumed $E_\nu\simeq E_{N_1}(M_2-M_1)/(4M_1)$, so the incident $N_1$ energies are $E_{N_1}=(3.2,32,320)$~EeV, respectively for $E_\nu\simeq(0.06,0.6,6)$~EeV. The assumed parameter values in benchmark (A) correspond to $\Gamma^{-1}_{tot}\simeq 42~{\rm km}\,(E_{N_1}/{\rm EeV})$ which is comparable with the $\nu_\tau$ mean free path length in the mantle $\tau_\nu\simeq800$~km at $E_{N_1}\simeq20$~EeV (or $E_\nu\simeq0.5$~EeV), and $\gamma^{-1} \simeq7,730$~km which is comparable to the chord size for $\theta_z\simeq130^\circ$.

\begin{table}[h!]
\caption{The benchmark values of the parameters of the model used in the Figures~\ref{fig:prob-z}, \ref{fig:prob-z-para} and \ref{fig:prob-e}. The second-to-last column shows the resulting decay length of $N_2$ particles  where $\tau_{N_2}$ is the lifetime of $N_2$ in its rest frame: $\Gamma_{tot}^{-1}=\tau_{N_2}\, (E_{N_1}/{\rm EeV})$. The last column is the mean free path of $N_1$ particles given by $\gamma^{-1}=(n_e\sigma)^{-1}$ for asymptotically large values of $s$ and assuming constant matter density with $n_e=2N_A~{\rm cm}^{-3}$ in the mantle ($N_A$ is the Avogadro number). For all the benchmarks $2m_\tau>M_2-M_1$, so the value of $B$ in Eq.~(\ref{eq:prob}) is $1/4$.} 
\begin{center}
\resizebox{\textwidth}{!}{\begin{tabular}{|c||*{7}{c|}}\hline
\centering
\backslashbox{\small Benchmark}{\small Parameter}
&\makebox[1.5em]{$g_N$}&\makebox[2.2em]{$g_{e-\tau}$}&\makebox[4.em]{$m_{Z'}$~[GeV]}
&\makebox[3.5em]{$M_2$~[GeV]}&\makebox[3.5em]{$M_1$~[GeV]}&\makebox[3.1em]{$\tau_{N_2}$~[km]}&\makebox[3.4em]{$\gamma^{-1}$~[km]}  \\\hline\hline
(A) & 3.0 & $6.9\times10^{-2}$ & 100 & 10 & 9.3 & 42 &7,730 \\\hline
(B) & 2.0 & $6.9\times10^{-2}$ & 100 & 10 & 9.3 & 94 & 17,390 \\\hline
(C) & 3.0 & $1.4\times10^{-1}$ & 200 & 10 & 9.3 & 168 & 7,730 \\\hline
(D) & 3.0 & $6.9\times10^{-2}$ & 100 & 10 & 8.7 & 2 & 7,730 \\\hline
\end{tabular}}
\end{center}
\label{tab:bench}
\end{table}%

Notice that in the above discussion, and in Figure~\ref{fig:prob-z}, we have assumed that a monochromatic neutrino will emerge from the Earth from the incident flux of monochromatic $N_1$ particles on the Earth. Obviously, this is only a proxy because both the $N_2$ produced in the interaction of $N_1$ with the electrons in Earth  and  the neutrinos produced in the $N_2$ decay have continuous spectra. Thus, in principle, one has to calculate the success rate $P$ for the spectrum of emerging neutrinos, for a fixed zenith angle, by taking into account the energy distributions. However, the energy, and also the zenith angle, dependence of the ANITA exposure is not available; which means that in order to calculate the number of events in ANITA one has to convolute the integrated success rate with the available (estimated) integrated exposure of ANITA. A shortcut to this procedure is to take the average energies $E_{N_2}\simeq E_{N_1}/2$ and $E_\nu\simeq E_{N_2}(M_2-M_1)/(2M_1)$ as has been adopted in Figure~\ref{fig:prob-z}.

As can be seen in Figure~\ref{fig:prob-z}, the expected number of events close to the horizon $\theta_z\lesssim95^\circ$ is small. The kink at $\theta_z\simeq 150^\circ$ corresponds to the chord tangent to the core; {\it i.e.,} for $\theta_z >150^\circ$ the $N_1$ crosses the core. The maximal success rate, $P\simeq 0.012$, occurs for $\theta_z\simeq(100^\circ-110^\circ)$. At the zenith angles of events observed by ANITA, shown by vertical dashed lines, the traversing probability reaches $\sim 0.01$. Assuming an exposure of $2.7~{\rm km}^{2}~{\rm yr}~{\rm sr}$ for ANITA~\cite{Schoorlemmer:2015afa}, this success rate requires an incident flux of $N_1$ particles on the Earth equal to $E_{N_1}\phi_{N_1}\simeq 40~{\rm km}^{-2}~{\rm yr}^{-1}~{\rm sr}^{-1}$ at $E_{N_1}\simeq 20$~EeV (the $\phi$ being the differential flux) to provide $\simeq1$ event at $E_\nu\simeq0.6$~EeV.\\

Notice that as we discussed in sect.~\ref{N1production}, in our scenario there is an accompanying flux of $\nu_e$ and $\nu_\tau$ along with the $N_1$ flux from the source. For both production mechanisms discussed in sect.~\ref{N1production}, the $N_1$ and $N_2$ are produced at the same rate. The subsequent decay of the $N_2$ particles within the source would create secondary $N_1$ particles with an energy almost equal to that of the parent $N_2$ energy (which has the same energy spectrum as that of the primary $N_1$ flux). Thus, effectively, the total flux of $N_1$ leaving the source is equal to the primary flux of $N_1$ plus the flux of $N_2$. The flux of neutrinos produced in the $N_2$ decay is $\sim1/4$ of the total $N_1$ flux (the total $N_1$ is two times the $N_2$ production flux, and the $N_2$ produces neutrinos in $1/2$ of the decays). The energy of the accompanying neutrinos to $N_1$ particles is $\mathcal{O}(E_{N_1} (M_2-M_1)/(2 M_1))$, or two times the Earth-emerging neutrinos from the later interaction and propagation of $N_1$ particles in the Earth. Taking into account these relative production rates, in all the range of energies of interest, the accompanying neutrino flux is compatible with the upper limits on diffuse neutrino flux. For example, the accompanying neutrino flux to the required $N_1$ flux for interpreting the ANITA events is $E_{\nu}\phi_{\nu}\simeq 10~{\rm km}^{-2}~{\rm yr}^{-1}~{\rm sr}^{-1}$ at $E_\nu\simeq1$~EeV, which is below the upper limit on the diffuse flux of neutrinos from Auger~\cite{Aab:2015kma} and is marginally compatible with the upper limit $E_{\nu}\phi_{\nu}\lesssim 7~{\rm km}^{-2}~{\rm yr}^{-1}~{\rm sr}^{-1}$ at $E_\nu\simeq1$~EeV from IceCube~\cite{Aartsen:2018vtx}.

From Figure~\ref{fig:prob-z}, one expects more events in ANITA at $\theta_{z}\sim100^\circ$ than the observed $\theta_z=117^\circ$ and $125^\circ$. However, two remarks are in order: \textit{i}) the difference between the maximal $P$ and the values of $P$ within the observed range ({\it i.e.,} between the vertical lines) is not large. In fact, it can be shown that the probability of observing the two registered events at the mentioned zenith angles while observing zero events in smaller zenith angles is $\sim50\%$; \textit{ii}) the number of background events in ANITA is larger near the horizon and so the sensitivity of detector decreases close to the horizon. At the same time,  decreasing $P$ near the horizon justifies the lack of events in the Auger experiment. The zenith coverage of Auger~\cite{Aab:2015kma} in searches for neutrino is $58.5^\circ<\theta_z<95^\circ$, where the value of $P$ is quite small.

\begin{figure}[t!]
\centering
\includegraphics[width=0.7\textwidth]{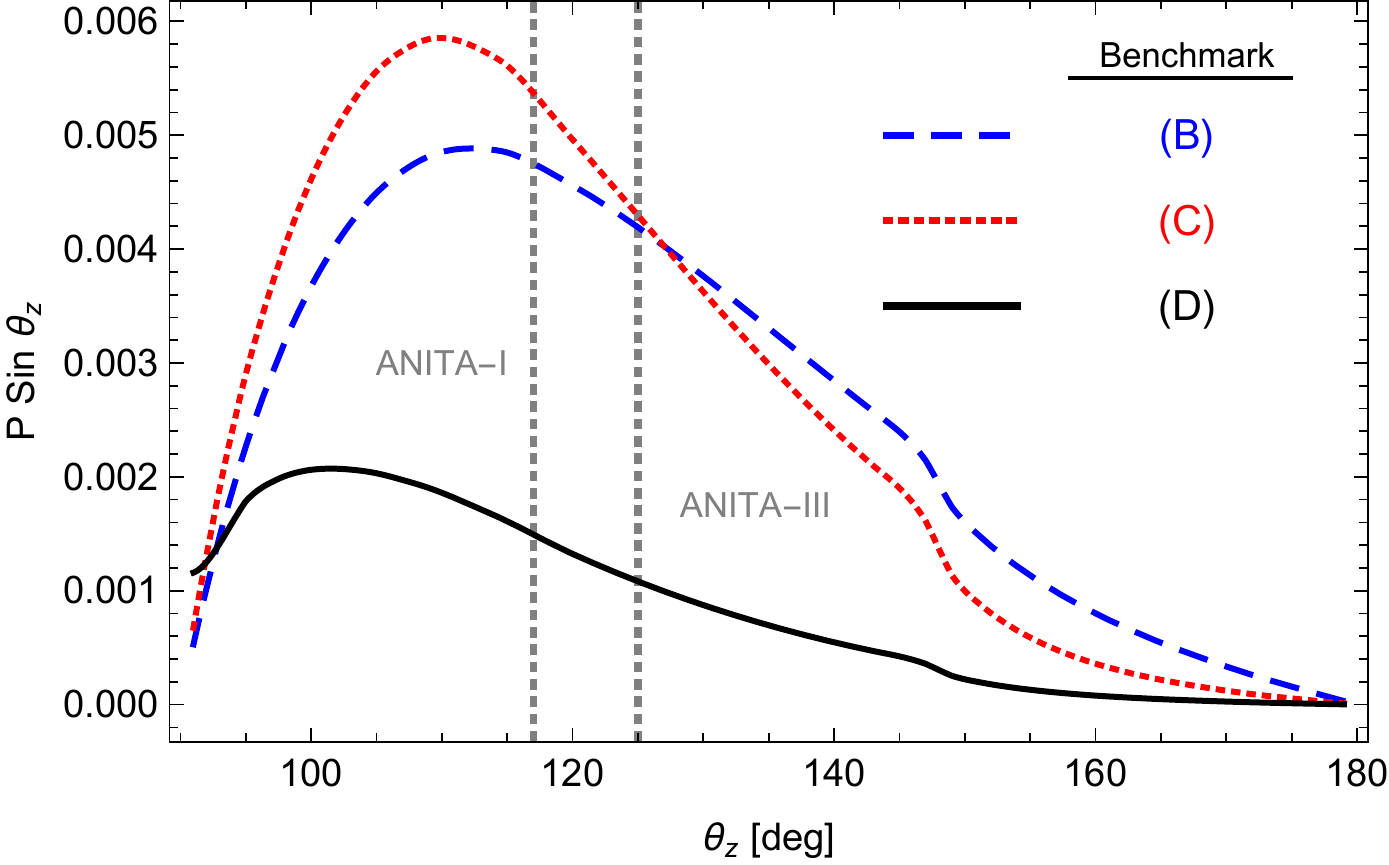}
\caption{\label{fig:prob-z-para} As in Figure~\ref{fig:prob-z}, the success rate $P$ as function of zenith angle $\theta_z$, for the various benchmark values of the parameters of model, labeled and shown in Table~\ref{tab:bench} (To be compared with the blue-dashed curve in Figure \ref{fig:prob-z} corresponding to benchmark (A)).}
\end{figure}

The zenith distribution of expected events, shown in Figure~\ref{fig:prob-z} and discussed above, qualitatively depends on the chosen benchmark values of the parameters in the model. Although we are not aiming for obtaining the preferred parameter space of the model for the interpretation of ANITA events, some general considerations can be outlined. Varying the masses of the $N_1$ and $N_2$ particles and choosing larger $M_2-M_1$, equivalent to larger $\Gamma_{tot}$ or smaller $N_2$ decay length $\Gamma_{tot}^{-1}$, the success rate $P$ decreases and the less-pronounced peak of the $P$ shifts to smaller zenith angles. For example, for $M_2=10$~GeV and $M_1=8.7$~GeV, the largest value of $P$ is $\sim2\times10^{-3}$ occurring at $\theta_z\sim100^\circ$. For this case, corresponding to the benchmark (D) in Table~\ref{tab:bench}, the $P$ dependence on $\theta_z$ is shown by the black-solid curve in Figure~\ref{fig:prob-z-para}. The flatter zenith distribution of $P$ for the black-solid curve is a consequence of the dependence of $\Gamma_{tot}$, and the independence of $\gamma$, on $M_1$ and $M_2$. The $\Gamma_{tot}$ and $\gamma$ have the same dependence on the couplings $g_N$ and $g_{e-\tau}$, and as far as the bounds in Eq.~(\ref{bAnds}) are satisfied, the increase in $g_N$ can be compensated by a decrease in $g_{e-\tau}$ and vice versa. Keeping  $g_{e-\tau}$ fixed, increasing (decreasing)  $g_N$ leads to an increase (decrease) of  $P$ (the same is true for fixing  $g_N$ and changing $g_{e-\tau}$). For example, in the benchmark (B) of Table~\ref{tab:bench} which is depicted by the blue-dashed curve in Figure~\ref{fig:prob-z-para}, with $g_N=2$,  $P$ scales down by a factor of $\sim3$. The $m_{Z'}$ dependence is milder in the region of interest. For example, setting $m_{Z'}=200$~GeV while  $g_{e-\tau}$ saturates the bound in Eq.~(\ref{bAnds}), the zenith distribution of $P$ scales down by a factor of $\sim2$ while the peak of $P$ shifts to larger zenith $\theta_z\simeq110^\circ$. This case, corresponding to the benchmark (C), is shown by the red-dotted curve in Figure~\ref{fig:prob-z-para}. Let us emphasize that although the success rate $P$ decreases for some choices of the parameters, as is the case for the three benchmarks (B,C,D) of Table~\ref{tab:bench} illustrated by the three curves in Figure~\ref{fig:prob-z-para}, these benchmarks may even better accommodate the ANITA observation. For example, the benchmark (C) corresponding to red-dotted curve in Figure~\ref{fig:prob-z-para} has a peak even at larger zenith angles compared to Figure~\ref{fig:prob-z}. The smaller probability means that for the interpretation of ANITA events a larger incident flux of $N_1$ on the Earth is required; which can lead to an accompanying neutrino flux larger than the upper limit of Auger~\cite{Aab:2015kma} and IceCube~\cite{Aartsen:2018vtx} on the diffuse flux of neutrinos. However, by considering the transient sources, which needs to be assumed as we discuss next, the upper limit on diffuse flux does not apply anymore.

\begin{figure}[t!]
\centering
\includegraphics[width=0.7\textwidth]{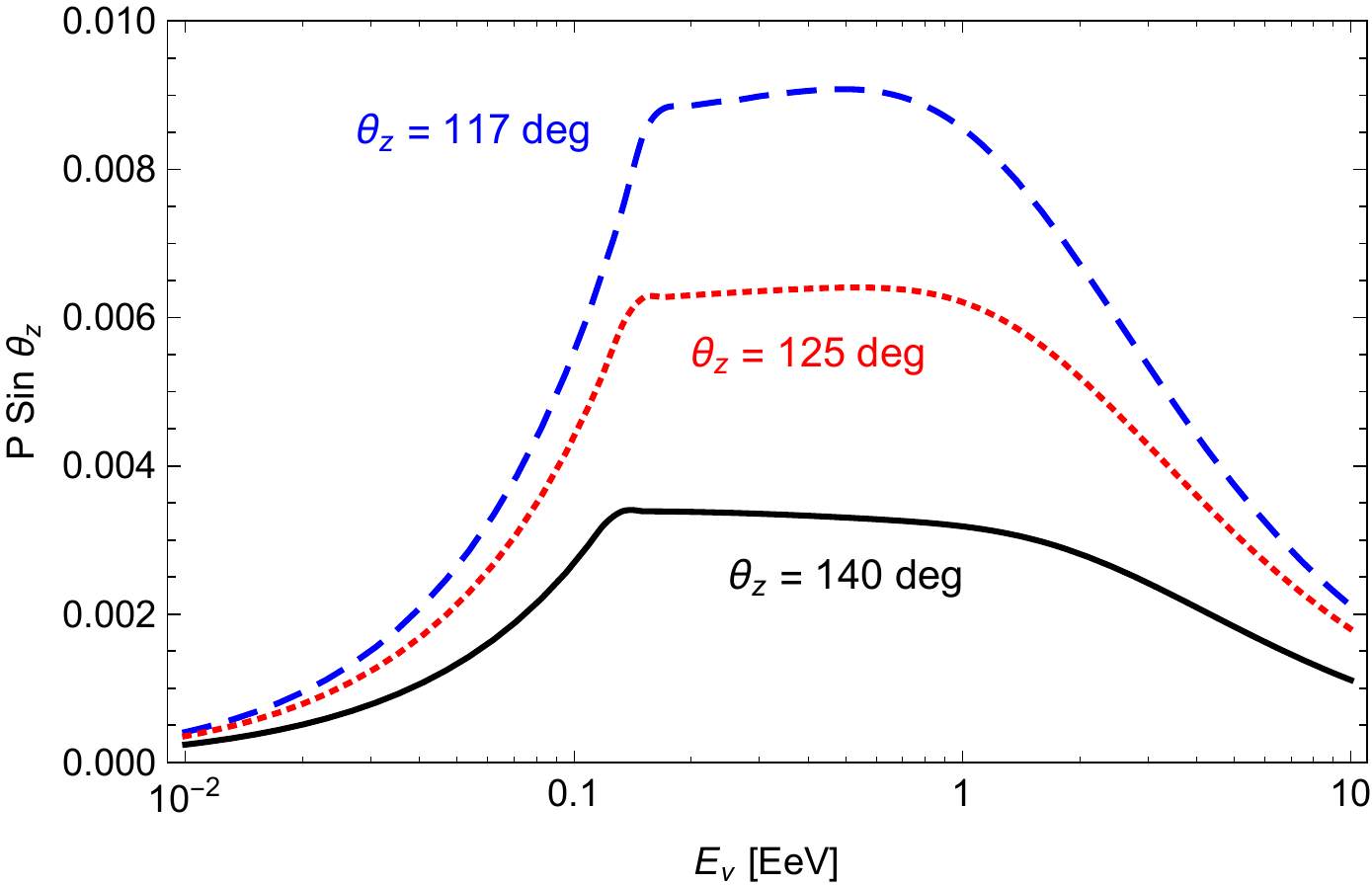}
\caption{\label{fig:prob-e} The energy dependence of $P$ in benchmark (A) for three different zenith angles $\theta_z=117^\circ,125^\circ,140^\circ$.}
\end{figure}

The energy dependence of the success rate $P$ in benchmark (A), for various zenith angles, is shown in Figure~\ref{fig:prob-e}. As in the previous figures, in Figure~\ref{fig:prob-e} also we use the approximation $E_\nu\simeq E_{N_1}(M_2-M_1)/(4M_1)$. Notice that the distributions in Figure~\ref{fig:prob-e}, after exchanging $E_\nu$ with $E_{N_1}$, have to be convoluted with the incident flux of $N_1$ and the exposure of ANITA to obtain the energy distribution of events, where the latter is not provided by the ANITA collaboration. The flux of $N_1$ generally has a power-law dependence on the energy, for a power-law energy distribution of accelerated particles in the source, or is monochromatic when the $N_1$ particles are produced in the dark matter decay. The $P$ dependence on the neutrino energy is flat in $E_\nu\sim(0.1-1)$~EeV, where the two ANITA events are located, and decreases in lower and higher energies. For a power-law flux of $N_1$ particles, the expected number of multi-EeV events will be much smaller than the sub-EeV events, for all the zenith angles, due to the decrease in both  $P$ and the incident flux of $N_1$. The expected number of sub-EeV events depends on the exact energy-dependence of the $N_1$ flux and it can be comparable to the number of events in $(0.1-1)$~EeV.

With the zenith and energy distributions of expected Earth-emerging neutrinos in our scenario in Figures~\ref{fig:prob-z} and \ref{fig:prob-e}, the lack of events in Auger and the observed values of zenith angle and energy in ANITA can be justified. However, the non-observation of the corresponding up-going events in IceCube experiment still requires a justification. Especially since the diffuse exposure of IceCube, estimated as $54~{\rm km}^{2}~{\rm yr}~{\rm sr}$ in~\cite{Fox:2018syq}, is one order of magnitude larger than ANITA, at least a factor of $\sim20$ larger number of events have to be detected in IceCube. However, the main contribution to the IceCube's large exposure is the data-taking time, $\sim8$ years compared to $\sim1$ month for ANITA, while the effective area of ANITA is much larger than the IceCube's effective area. In~\cite{Cherry:2018rxj} the transient acceptances of ANITA and IceCube have been calculated and compared, which shows comparable values at $\sim$~EeV. Thus, reconciling the ANITA and IceCube in our model requires the assumption of transient sources, as in any other model proposed up to now. Assuming the transient sources, the non-observation of events in IceCube in $\lesssim1$~EeV has a chance of $50\%$~\cite{Cherry:2018rxj}.

Based on Figures~\ref{fig:prob-z} and \ref{fig:prob-e}, we can qualitatively predict the future observations in ANITA and IceCube in our scenario. Assuming the transient sources, the rate of observation depends on the rate of transients and the angular distribution of transients. Having two events in ANITA it is not possible to quantify the rate of transients, although one can assume a rate of $\sim 1~{\rm yr}^{-1}$. Notice that the rate can be very different especially since the data-taking periods of ANITA are very short and happen once per few years and so no decisive conclusion can be drawn based on the ANITA observation of two events in 2006 and 2014. For the moment let us assume that the angular distribution of the transients is isotropic. For IceCube, with $\sim8$ years of data-taking, each transient if it happens at $\theta_z\sim100^\circ-130^\circ$ can lead to $\sim1$ event (see also the figure 3 in~\cite{Cherry:2018rxj}). At larger zenith angles the expected number of events in IceCube is smaller based on Figure~\ref{fig:prob-z}. Of course, a non-isotropic distribution of transient sources can lead to larger or smaller number of events in IceCube depending on the distribution. Improvement in the background rejection in ANITA near the horizon can lead to observation of events in $\theta_z\simeq 100^\circ$, although the exact expectation depends on the background rejection level. We expect to observe the future events mainly in the energy range $E_\nu\sim(0.1-1)$~EeV, while the lower energy $E_\nu\lesssim0.1$~EeV will be populated especially for $\theta_z\sim100^\circ-120^\circ$ and for steep flux of $N_1$ extended to lower energies.

\section{Concluding remarks}\label{Sum}

We have proposed a model for explaining the two anomalous neutrino events observed by ANITA. The model adds two Weyl fermions $N_1$ and $N_2$ with a naturally small mass splitting. These two fermions couple to a new gauge boson $Z'$ that converts them to each other. A $Z_2$ symmetry stabilizes the lighter one, $N_1$. The gauge boson $Z'$ couples to the first and the third generations of leptons through a $L_e-L_\tau$ gauge symmetry. A flux of $N_1$, produced via the $e^- e^-$ interaction at a far away source or from dark matter decay, arrives at the Earth, then interacts via the new gauge symmetry with the electrons inside the Earth and converts into $N_2$. Subsequently, $N_2$ decays into $N_1$ and a pair of leptons with $1/3$  (or $1/4$ depending on whether $M_2-M_1> 2m_\tau$ or $M_2-M_1<2m_\tau$) of chance of producing $\nu_\tau$ and $\bar{\nu}_\tau$. The energies of final $\nu_\tau$ and $\bar \nu_\tau$ will be about  2-3~\% of the initial energy of $N_1$. In summary, if the value of the gauge coupling of $Z'$ to the electron saturates the bound from the LEP, the probability of the initial $N_1$ leading to the emergence of a $\nu_\tau$ or $\bar{\nu}_\tau$ in the vicinity of the Earth surface is $\mathcal{O}(1\%)$. This may be compared to the suppression of $10^{-7}-10^{-6}$ for the flux of standard $\nu_\tau$ and $\bar \nu_\tau$ of similar energy crossing the Earth. 
 
The dependence of the probability on the zenith angle is shown in Figure~\ref{fig:prob-z}. The suppression of the probability close to the horizon ({\it i.e.,} $\theta_z=90^\circ$) explains the bounds from Auger. Moreover the lack of events with zenith angles different than what is observed can also be explained. However, considering the much larger exposure of IceCube compared to that of ANITA, the lack of events observed by IceCube requires some explanation. Like~\cite{Cherry:2018rxj}, we may argue that the source were transients to explain the tension; however, the occurrence of such transients do not need to be very rare. Any transient occurring at the $\theta_z\sim100^\circ-130^\circ$ would lead to $\sim1$ events in IceCube. At larger zenith angles the expected number of events decreases. 

Dependence of the predictions of the proposed model on the assumed benchmark for parameters has been demonstrated in Figure~\ref{fig:prob-z-para} (see Table~\ref{tab:bench} for various benchmarks). For various benchmarks of the parameters the model retains qualitatively the desired zenith-dependence of $P$. The energy spectrum of emerging neutrinos is almost flat in the range $E_\nu\sim(0.1-1)$~EeV (see Figure~\ref{fig:prob-e}). At $E_\nu\gtrsim1$~EeV the model predicts a small success rate for emerging neutrinos; while for $E_\nu\lesssim0.1$~EeV, depending on the energy dependence of the $N_1$ flux, future observation of events is expected (mainly in IceCube since ANITA loses sensitivity for $\lesssim0.1$~EeV).      

In our model, $Z'$ does not couple to quarks so the LHC cannot probe it but at the ILC, $Z'$ can be produced via $e^-e^+ \to \gamma Z'$. We have discussed the distinctive signatures of $Z'$ at the ILC and have suggested strategies to determine the parameters of our model. If ILC does not find any $Z'$ with characteristics described in this paper, a stronger bound will be set on the gauge coupling, reducing the probability of $N_1$ interacting in Earth and leading to $\nu_\tau$ or $\bar{\nu}_\tau$ events at ANITA. This in turn means that to explain the ANITA event, we need a higher flux of $N_1$ and therefore a more powerful source.

The $N_1$ particles in our model are stable and electrically neutral and can contribute to the dark matter content of the Universe. We show that the co-annihilation with $N_2$ sets the abundance, preventing the overclosure of the Universe.


\acknowledgments
This project has received funding from the European Union\'~\!s Horizon 2020 research and innovation programme under the Marie Sklodowska-Curie grant agreement No.~674896 and No.~690575. YF has received partial financial support from Saramadan under contract No.~ISEF/M/98223. YF would like also to thank the  ICTP staff and the INFN node of the INVISIBLES network in Padova. AE thanks the partial support by the CNPq fellowship No.~310052/2016-5. AE would like to thank ICTP where this project started.


\end{document}